\documentclass[manuscript]{acmart}
\usepackage{lipsum} 
\usepackage{hyperref}
\AtBeginDocument{%
  \providecommand\BibTeX{{%
    \normalfont B\kern-0.5em{\scshape i\kern-0.25em b}\kern-0.8em\TeX}}}

\setcopyright{none}
\renewcommand\footnotetextcopyrightpermission[1]{}
\begin{document}

\title[Simulation as Experiment]{Simulation as Experiment: An Empirical Critique of Simulation Research on Recommender Systems}

\author{Amy A. Winecoff}%, Matthew Sun, Eli Lucherini, Arvind Narayanan}
\affiliation{%
  \institution{Princeton University}
    \city{Princeton}
  \state{New Jersey}
  \country{USA}
  }
 \email{aw0934@princeton.edu}

\author{Matthew Sun}
\affiliation{%
  \institution{Princeton University}
    \city{Princeton}
  \state{New Jersey}
  \country{USA}}
\email{mdsun@princeton.edu}

\author{Eli Lucherini}
\affiliation{%
  \institution{Princeton University}
  \city{Princeton}
  \state{New Jersey}
  \country{USA}}
\email {elucherinin@cs.princeton.edu}

\author{Arvind Narayanan}
\affiliation{%
 \institution{Princeton University}
   \city{Princeton}
  \state{New Jersey}
  \country{USA}}
\email{arvindn@cs.princeton.edu}

\renewcommand{\shortauthors}{Winecoff, et al.}

% <concept_id>10010147.10010341</concept_id>
% <concept_desc>Computing methodologies~Modeling and simulation</concept_desc>

% \ccsdesc[500]{Computing methodologies~Modeling and simulation}

% <concept_id>10010147.10010341</concept_id>
% <concept_desc>Computing methodologies~Modeling and simulation</concept_desc>
% \ccsdesc[500]{Computing methodologies~Modeling and simulation}

% %% the work being presented. Separate the keywords with commas.
% \keywords{recommender systems, simulation, research methods, replication}

\maketitle
\section{Introduction}

Simulation can enable the study of recommender system (RS) evolution while circumventing many of the issues of empirical longitudinal studies; simulations are comparatively easier to implement, are highly controlled, and pose no ethical risk to human participants. How simulation can best contribute to scientific insight about RS alongside qualitative and quantitative empirical approaches is an open question. Philosophers and researchers have long debated the epistemological nature of simulation compared to wholly theoretical or empirical methods \cite{parker2008franklin, parker2009does, parker2020evidence, reiss2011plea, winsberg2003simulated, peck2004simulation, davis2007developing}. Simulation is often implicitly or explicitly conceptualized as occupying a middle ground between empirical and theoretical approaches, allowing researchers to realize the benefits of both \cite{davis2007developing, winsberg2003simulated}. However, what is often ignored in such arguments is that without firm grounding in any single methodological tradition, simulation studies have no agreed upon scientific norms or standards, resulting in a patchwork of theoretical motivations, approaches, and implementations that are difficult to reconcile.

In this position paper, we argue that simulation studies of RS are conceptually similar to empirical experimental approaches and therefore can be evaluated using the standards of empirical research methods. Using this empirical lens, we argue that the combination of high heterogeneity in approaches and low transparency in methods in simulation studies of RS has limited their interpretability, generalizability, and replicability. We contend that by adopting standards and practices common in empirical disciplines, simulation researchers can mitigate many of these weaknesses.

\section{Background \& Motivation}

In simulation, the researcher develops a model system to study a target system. Sometimes the target system is a real-world phenomena; the researcher conducts experiments in the model system to draw inferences about its real-world analog. For example in the 1950s, the Army Corps of Engineers developed a physical, functioning model of the San Francisco Bay to determine whether constructing a dam would improve the stability of the city’s water supply. The model leveraged hydraulic pumps designed to mimic water flow patterns in the Bay, which were carefully calibrated using measurement collected in the Bay itself. Simulations using the model Bay suggested that the proposed dam would not improve the water supply but would have disastrous downstream ecological consequences \cite{weisberg2012simulation, armycorps}. 
 
Although the Bay model was a simplified system, it was used to evaluate a potential intervention for a real-world target system. Where the goal of a simulation is to draw inferences about real-world processes or real-world interventions, reproducing the relevant real-world conditions is critical. However, the goal of many simulations is not to reproduce the specifics of real-world phenomena, but rather to demonstrate a theory about how system-level outcomes emerge. In these cases, simulations are similar to empirical laboratory experiments, which eschew the complexities of the real world in order to characterize the theoretical causal relationship between variables that are often difficult to cleanly isolate in real-world settings [1]. Unlike the Bay model, the target system for many simulation models is an experiment. 

When the target system for a simulation is an experiment, maximizing experimental control is critical. Intuitively, it would seem important for an experimental design to mimic aspects of real-world circumstances, thereby bolstering ecological validity\footnote{Ecological validity is defined as the extent to which the experimental environment mirrors the real-world environment \cite{stangor2014research}}; however, misguided efforts to attain ecological validity can undermine the very point of an experiment--to cleanly attribute changes in an outcome to an experimental manipulation \cite{banaji1989bankruptcy, mook1983defense, reiss2019against}. Schelling’s \cite{schelling1971dynamic} computational model of neighborhood segregation is an example of an ecologically-invalid setup that nevertheless effectively demonstrates a theoretical relationship under tightly controlled circumstances. Very little about the simulation setup reflects the real-world conditions in which neighborhood segregation occurs, but the study's power lies in its simplicity; a single agent-level preference is sufficient to cause system-level patterns of segregation that are qualitatively similar to patterns observed in real-world neighborhoods, even if they do not mimic them exactly. As with a well-designed experiment, the Schelling simulation demonstrated one simple causal mechanism (homophily) through which segregation could occur. 

The standards and practices of empirical research methods were developed to help researchers realize their primary goal--to validly and reliably evaluate the relationship between variables. When simulation studies share the same primary goal, the standards of experimental research can be used to evaluate such studies' methodological rigor and scientific merit. Yet, in aggregate much of the extant literature using simulation to study RS would fail to meet these standards. In the next section, we describe several such failures and why they are cause for concern.

\section{High Heterogeneity and Low Transparency}
Before the relationship between variables can be assessed, individual variables must first be defined or “operationalized” into quantifiable terms. For example, depending on the researcher’s specific question, she might operationalize the conceptual variable of “user price sensitivity” in a variety of ways. If she is primarily interested in users’ self-perceptions of price sensitivity, she might employ a survey designed to measure the effect of price on consumers' product evaluations. On the other hand, if the researcher is more interested in behavioral indicators of price sensitivity, she might measure users’ price sensitivity implicitly through a decision making task that involves weighing product price against other product attributes. To assess the causal relationship between two variables experimentally, researchers manipulate the hypothesized causal variable (i.e., independent variable) and measure the impact on the outcome variable (i.e., dependent variable). Likewise, simulation studies must operationalize theoretical constructs mathematically and determine which variables' manipulations will best demonstrate the causal relationship between them. 

Early in the development of a discipline, how researchers operationalize theoretical constructs is likely to be heterogeneous. This heterogeneity is not necessarily a problem. By providing multiple definitions for related constructs, researchers can examine different aspects of the same concept, which collectively provide a more comprehensive understanding of a complex phenomenon. Within the social sciences, the capacity of any measure or instrument to adequately capture all facets of a phenomenon is referred to as “content validity” \cite{cronbach1955construct}. Measures with low content validity can lead to an incomplete or inaccurate understanding of a multifaceted construct. For example, if a test for humans’ general reasoning abilities only evaluated spatial reasoning, the test would be low in content validity since general reasoning also involves verbal, mathematical, and social components. Yet a proliferation of operational definitions for the same construct can also lead to conceptual confusion and can preclude any meaningful generalizations about the overall phenomena; if we do not know how two different definitions of the same concept are related within a single study, we certainly cannot make any inferences about how they are related between studies. 

Several recent studies \cite{chaney2018, jiang2019degenerate, aridor2020deconstructing, mansoury2020feedback} have used different simulation designs to study "filter bubbles" in RS, whereby users are exposed to an increasingly narrow range of content that matches their existing preferences, further homogenizing their interests \cite{pariser2011filter}. The simulations in Chaney et al. \cite{chaney2018} and Mansoury et al. \cite{mansoury2020feedback} suggest that algorithmic feedback loops can over-homogenize user preferences \cite{chaney2018} and exacerbate algorithmic bias towards popular items and away from the minority subpopulation's interest \cite{mansoury2020feedback}. In contrast, the simulations in Jiang et al. \cite{jiang2019degenerate} and Aridor et al. \cite{aridor2020deconstructing} locate the cause of filter bubbles in the preferences and behavior of users and suggest that RS can either alter the rate at which users’ interest homogenize \cite{jiang2019degenerate} or prevent the homogenization of users’ interest that would occur in the absence of recommendation \cite{aridor2020deconstructing}. Both Jiang et al. \cite{jiang2019degenerate} and Aridor et al. \cite{aridor2020deconstructing} centrally incorporate some notion of user preference bias. In Jiang et al. \cite{jiang2019degenerate}, users’ preference for an item increases every time the user interacts with that item. In Aridor et al.,  \cite{aridor2020deconstructing}, users vary in risk aversion and how much their beliefs about other items are shifted based on item consumption.

A casual reader might conclude that the user preference model in Jiang et al. \cite{jiang2019degenerate} and Aridor et al. \cite{aridor2020deconstructing} could explain why these two simulations suggest a different mechanism for the emergence of filter bubbles than in Chaney et al.  \cite{chaney2018} and Mansoury et al. \cite{mansoury2020feedback}, where no user biases are incorporated. Yet numerous other sources of heterogeneity in these four simulations’ theoretical constructs and their specific operationalizations preclude any inference about why they have arrived at seemingly contradictory conclusions. For example, the entities over which homogenization is calculated differs across these studies. Aridor et al. \cite{aridor2020deconstructing} primarily define homogenization as the distance in a stylized space between the items users interact with between adjacent timepoints (i.e., within-user homogenization). Jiang et al. \cite{jiang2019degenerate} use a different within-user, across time definition of homogenization. In contrast, Chaney et al. \cite{chaney2018} define homogenization as the overlap between sets of items consumed by pairs of users (i.e., between-user homogenization). Mansoury et al. \cite{mansoury2020feedback} define homogenization based on preference distribution shifts both within user using the Kullback-Leibler divergence (KLD) between a user's initial and current preferences, as well as between groups of users as the KLD between male and female subpopulations of users (i.e., between-groups homogenization). Because these studies’ definitions of homogeneity differ, the proposed causal mechanisms for filter bubble effects cannot be readily compared. 

Notably, Aridor et al. \cite{aridor2020deconstructing} does include an analysis of between-user homogeneity using a similar operational metric to Chaney et al. \cite{chaney2018}. Like Chaney et al. \cite{chaney2018}, they also find that recommendation exacerbates filter bubble effects between users. Thus, it is possible that recommendation can simultaneously prevent within-user homogenization while perpetuating between-user homogeneity. The use of a shared metric allows for a comparison between the pattern of results for the two different simulation set ups. However, even with the single shared metric, it is unclear if or how the rest of the results can be synthesized since the two simulations differed in a range of other ways including how items are defined, which rules govern user behavior, and how recommendation is enacted. A justification for the many different operationalizations between studies is that each study addressed nuanced, but distinct questions and therefore required different specifications. Even so, it is nevertheless the case that another researcher seeking to synthesize these studies' results either could not do so because of the heterogeneity between studies or worse, might attempt to do so anyway and arrive at an erroneous conclusion.

% the numerous other differences in the simulation designs of these studies precludes synthesis of their findings even though they do share one operational definition of the system outcome variable. Synthesis of findings across all four studies even more tenuous. The authors of each of these studies could reasonably object that their own simulations addressed nuanced, but distinct questions and therefore required different operationalizations of the relevant constructs. While this point is valid, it is nevertheless the case that another researcher seeking to synthesize results from simulation studies of filter bubbles in RS either could not do so because of the heterogeneity between studies or worse, might attempt to do so anyway and arrive at an erroneous conclusion. 

% One justification for the many different operationalizations is that each study addressed nuanced, but distinct questions and therefore required different specifications. Even so, it is nevertheless the case that another researcher seeking to synthesize these studies' results either could not do so because of the heterogeneity between studies or worse, might attempt to do so anyway and arrive at an erroneous conclusion. 

In the social sciences, measurement instruments such as surveys typically undergo rigorous evaluation before they can be deployed \cite{fowler2013survey, bradburn2004asking}; thus, researchers are strongly encouraged to use existing measures instead of creating new ones \cite{salganik2019bit, bradburn2004asking}.  Analogously, in empirical studies of RS, individual researchers do not define their own notion of algorithm accuracy. The field has converged upon a handful of accepted accuracy metrics, which has facilitated scientific study of RS, particularly early in the discipline's development \cite{jannach2016recommendations}. Simulation researchers would similarly benefit from establishing several core metrics that are included in most simulation studies of RS. Even in cases where a novel metric is justified, including the core established metrics would facilitate conceptual alignment of the field and enable researchers to more easily compare results from related studies. In sum, if simulation studies of RS fail to converge on any common conceptualizations of variables and their operational definitions, the body of simulation studies as a whole contributes little generalizable scientific knowledge, even if individual studies are thought-provoking.

The problems associated with highly heterogeneous approaches are further magnified by methodological opacity. Where approaches are standardized, researchers do not need to painstakingly present every detail of their methods. For example, a social scientist that uses R's \texttt{lme4} package \cite{bates2007lme4} for generalized linear mixed models does not need to precisely describe the optimization procedure since details are available in the package documentation and source code. On the other hand, where tools are not standardized, much more methodological transparency is necessary. Motivated by failures to replicate seminal findings in fields such psychology \cite{open2015estimating} and cancer biology \cite{baker2017cancer}, empirical research communities have begun embracing an open science approach (see \url{https://www.cos.io/}). Among other goals, open science seeks to improve methodological accountability and transparency, which facilitate more replicable research. Empirical RS researchers have also begun embracing open science principles by pushing for the adoption of: 1) research best practices that improve research transparency and replicability \cite{konstan2013toward}; and 2) standardized implementations of commonly used recommendation algorithms \cite{ekstrand2011rethinking}. Empirical studies have demonstrated that a variety of user and contextual factors can undermine the replicability of results \cite{beel2016towards} and that widely touted improvements to recommendation algorithms are not robust when compared to strong baselines \cite{dacrema2019we}. These findings underscore the ongoing need for a more open and transparent approach to empirical research in general and on RS, specifically. 

The results of empirical studies can be influenced by a variety of factors that are outside of the researcher’s control. Each idiosyncrasy in how an empirical study was conducted presents a new avenue through which a replication effort could fail. In contrast, in simulation studies, researchers wield complete control over the design and implementation. As a result, it would seem at first blush that efforts to replicate findings from simulation studies would be infinitely less complex than efforts to replicate empirical findings. However, it is precisely because simulation researchers have so many degrees of freedom that simulation research is difficult to replicate. In simulation studies of RS, research groups frequently implement their simulation designs through ad-hoc codebases that are not open source. As a result, these simulation implementations differ in numerous non-obvious ways that cannot be resolved.  

In theory, the methods sections of published reports should be sufficient for an independent research group to replicate the original findings. In practice, replication of simulated results is less straightforward. As part of an effort to develop a standarized platform for conducting simulations of algorithmic systems \cite{lucherini2021trecs}, our research group recently replicated results from two simulations \cite{chaney2018, goel2016structural}. Both reports provided highly detailed information regarding the simulation methodology. However, replicating the core findings required months of effort from our team and a substantial amount of trial and error. In the trial and error process, we also discovered that differences in seemingly minor implementation choices often resulted in surprisingly large differences in the system outcomes. 

For example, Chaney et al., \cite{chaney2018} investigated whether algorithmic confounding in RS homogenizes users to a greater degree than an omniscient ``ideal'' recommendation algorithm, which always recommends items to users in line with their true preferences. They define the similarity of two users' \textit{behavior} at timestep $t$ as the Jaccard similarity between the sets of items that each user has interacted with. They then measure homogenization by averaging this Jaccard similarity over pairs of users, where each user is paired with the user with the most similar \textit{preferences}. They then normalize this measure relative to the ideal recommender to demonstrate how much each algorithm homogenizes user behavior over and above what would be ideal. 

\begin{figure}[!t]
\includegraphics[width=130mm,scale=0.5]{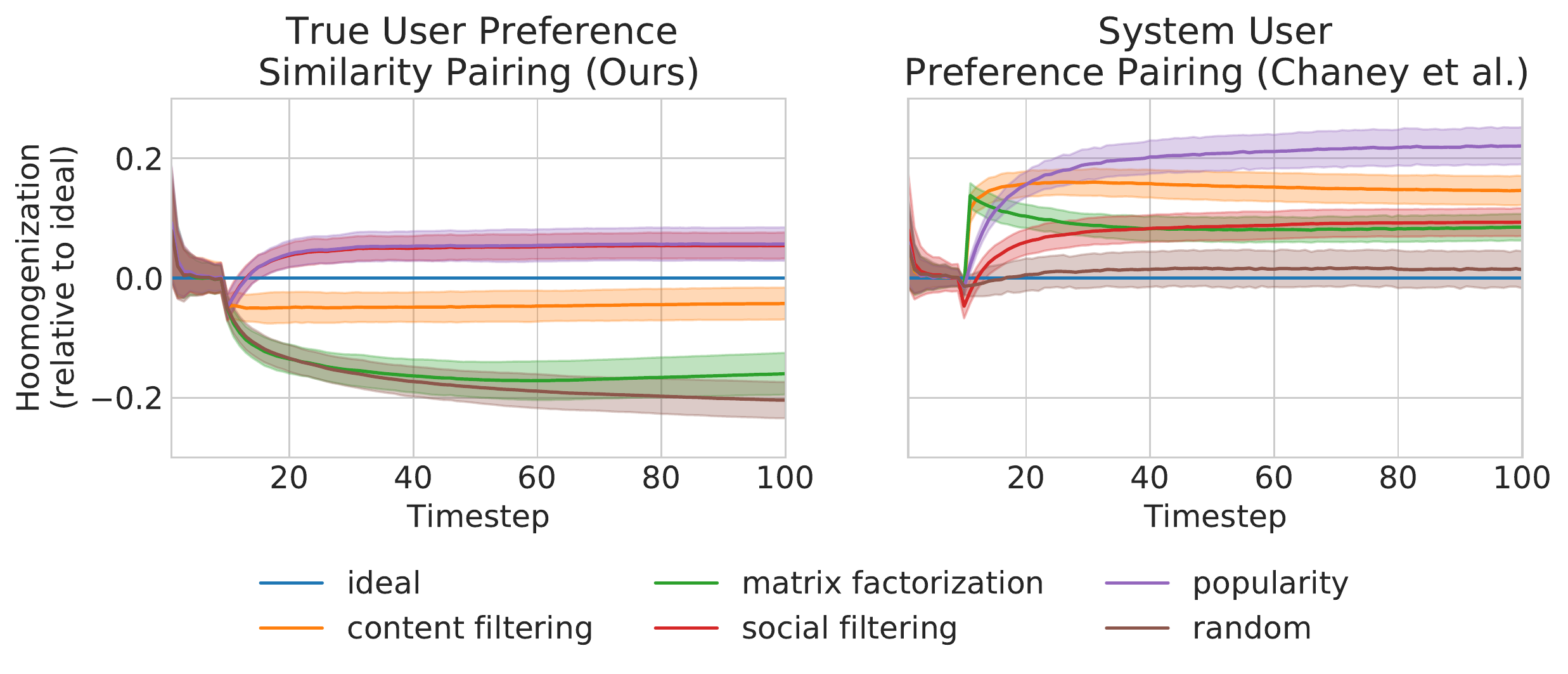}
\caption{Homogenization of user behavior relative to ideal for different algorithms. 10 initial timesteps of random recommendations are followed by 90 timesteps of recommendations, with retraining between timesteps. Left: Results using our definition of homogenization. Right: Results using Chaney et al.'s \cite{chaney2018} definition. Results averaged over 400 trials. The shaded region indicates $\pm$ 1 SD.}
\label{fig:compare_jacc_def}
\end{figure}

An important detail in Chaney et al.'s \cite{chaney2018} definition of homogenization is that they construct the pairs of users based on the algorithm's internal representation of user preferences, which changes each time the algorithm is retrained, rather than the users' true preferences, which are stable over time. Early in our replication attempt, we incorrectly assumed that users should be paired based on true preferences since that would result in a stable set of similar user pairs over the duration of the simulation and would allow a direct comparison between algorithms since pairs would also be the same for all algorithms. Both Chaney et al.'s definition of homogenization and our definition of homogenization are both reasonable ways to operationalize the phenomenon of homogenization. As Chaney et al. \cite{chaney2018} point out, ``Recommendation algorithms encourage similar users to interact with the same set of items, therefore homogenizing their behavior, relative to the same platform without recommended content.'' However, the slight change in operationalizations, which only differ by one line out of the hundreds of lines of code in our replication, result in starkly different results and qualitative conclusions about homogenization (See Figure \ref{fig:compare_jacc_def}). Our metric suggests that the matrix factorization and content filtering algorithms homogenize user behavior less than the ideal algorithm, and that the ideal algorithm homogenizes user behavior far more than the random recommender system. In contrast, Chaney et al.'s metric suggests that all recommender systems homogenize user behavior to a greater degree than the ideal recommender system. Thus, even holding a simulation environment largely constant, subtly different operationalizations provide very different information about the homogenizing effects of recommendation algorithms. 

\section{Conclusions}
The lack of consensus on best practices in simulation studies of RS has limited their potential scientific impact. Because approaches in this subdiscipline are neither standardized nor sufficiently transparent, simulation studies of RS cannot meaningfully inform one another, much less other RS using different methods. As in empirical experimental research, common definitions of key concepts, standardized implementations, and an embrace of open science practices can improve the robustness of simulation studies and the speed of innovation. Furthermore, by creating alignment in both theory and practice between empirical and simulation studies of RS, simulation researchers can improve the likelihood that their simulation research will have practical impact.

\begin{acks}
This work is partly supported by the 2021 Twitch Research Fellowship and the 2021-22 Hamid Biglari *87 Behavioral Science Fellowship.
We are also grateful for support from the National Science Foundation under Award IIS-1763642. Additionally, we gratefully acknowledge financial support from the Schmidt DataX Fund at Princeton University made possible through a major gift from the Schmidt Futures Foundation.
\end{acks}

\bibliographystyle{ACM-Reference-Format}
\bibliography{draft}
\end{document}